\documentclass[aps,pre,a4paper,twocolumn,groupedaddress,showpacs,showkeys,preprintnumbers,floatfix,dvips]{revtex4}

\usepackage[utf8]{inputenc} \usepackage[T1]{fontenc}
\usepackage{psfrag} 
\usepackage{times} \usepackage{graphicx} \usepackage{color}
\usepackage{amsmath} \usepackage{amscd} \usepackage{amssymb} \usepackage{amsthm}
\usepackage{ragged2e} \usepackage{subfigure} \usepackage{caption}
\usepackage{nicefrac}
\usepackage{captcont}

\DeclareCaptionJustification{reallyjustified}{\justifying}
\begin{document}

\title{On the exit probability of the one dimensional $q$-voter model. Analytical results and simulations for large networks.}
\author{Andr\'e M. Timpanaro} \email[]{timpa@if.usp.br} \author{Carmen P. C.
Prado} \email[]{prado@if.usp.br} \affiliation{Instituto de F\'{i}sica,
Universidade de S\~{a}o Paulo \\ Caixa Postal 66318, 05314-970 - S\~{a}o Paulo -
S\~{a}o Paulo - Brazil} \date{\today}


\graphicspath{ {./figuras_eps/} }


\begin{abstract}
We discuss the exit probability of the one dimensional $q$-voter model and present tools to obtain estimates about this probability both through simulations in large networks (around $10^7$ sites) and analyticaly in the limit where the network is infinetely large. We argue that the result $E(\rho) = \frac{\rho^q}{\rho^q + (1-\rho)^q}$, that was found in 3 previous works (2008 EPL 82 18006 and 2008 EPL 82 18007, for the case $q=2$ and 2011 PRE 84 031117, for $q>2$) using small networks (around $10^3$ sites), is a good approximation, but there are noticeable deviations for larger system sizes. We also show that, under some simple and intuitive hypothesis, the exit probability must obey the inequality, $\frac{\rho^q}{\rho^q + (1-\rho)} \leq E(\rho) \leq \frac{\rho}{\rho + (1-\rho)^q}$, in the infinite size limit. We believe this settles in the negative the suggestion made (2011 EPL 95 48005) that this result would be a finite size effect, with the exit probability actualy being a step function. We also show how the result, that the exit probability cannot be a step function, can be reconciled with the Galam unified frame, which was also a source of controversy.
\end{abstract}

\maketitle

\section{Introduction}

The non-linear $q$-voter model is an opinion propagation model, proposed by Castellano et al. \cite{q-votante-def} as a variant of the voter model. In this model, a society is represented by a network, where sites represent agents and the edges represent social connections among them. At each time step, an agent consults a group with $q$ neighbouring agents about some subject. If all the agents in the group agree with each other, they convince the first agent. In the original version of the model, if the agents in the chosen group don't agree, there is a probability $\epsilon$ that the first agent changes its opinion. In the works about the model that we will consider and with which we compare our results, this part of the model is ignored (that is, $\epsilon$ is set to 0). For this reason we will drop this rule, so in this work if the agents in the chosen group don't agree with each other, nothing happens and we go to the following time step. If we consider a one dimensional lattice, as the model of our society, then there is no difference between inflow and outflow dynamics and the model can be regarded as a generalization of sorts of the well known Sznajd model \cite{sznajd-def}.

Slanina, Sznajd-Weron and Przyby\l a in \cite{q-voter-Slanina}, and simultaneously Lambiotte and Redner in \cite{q-voter-Lambiotte}, studied the case $q=2$ (that corresponds to the Sznajd model \cite{sznajd-def}) in one dimension and with two opinions. The main conclusion is that the exit probability $E(\rho)$ (the probability that an opinion that starts with a proportion $\rho$ of the agents, ends up being the dominant opinion, with all agents adopting it) is a continuous function, given by

\begin{equation}
E(\rho) = \frac{\rho^2}{\rho^2 + (1-\rho)^2}.
\label{eq:q-2-continua}
\end{equation}
In \cite{q-voter-Galam}, Galam and Martins questioned the analytical arguments in the deduction of this exit probability (that are based in the Kirkwood approximation, a type of mean field treatment, as correlations beyond nearest neighbours are truncated \cite{kirkwood-approximation, q-voter-Slanina}). They put forward the idea that the exit probability could be a step function (as would be expected from the application of the Galam unifying frame (GUF) \cite{Galam-GUF}), and that equation \ref{eq:q-2-continua} is a consequence of finite size effects. Alternatively, they suggest that this could be an indication of the irrelevance of fluctuations in this system (because it can be derived from a mean field treatment).

In \cite{q-voter-Sznajd}, Przyby\l a, Sznad-Weron and Tabiszewski showed that something similar happens for a linear chain with $q>2$, when we forbid repetitions among the $q$ consecutive neighbours that are chosen (which is allowed in the original $q$-voter model but that makes no difference when $q=2$). The exit probability that was found is

\begin{equation}
E(\rho) = \frac{\rho^q}{\rho^q + (1-\rho)^q}.
\label{eq:q-continua}
\end{equation}
Once more, applying the GUF yields a step function and, for this case, there is still no derivation for equation \ref{eq:q-continua}.

In this work, we define a dual model that is mathematicaly equivalent to the $q$-voter model studied in \cite{q-voter-Sznajd} (section \ref{ssec:dual}). We use it to show a connection between this model and the usual voter model. This connection allows us to make estimates about the exit probability and we are able to derive, through analytical arguments and rather intuitive hypothesis, the inequality

\begin{equation}
\frac{\rho^q}{\rho^q + (1-\rho)} \leq E(\rho) \leq \frac{\rho}{\rho + (1-\rho)^q}
\label{eq:bounds}
\end{equation}
for the limit of an infinite system size, ruling out a step function as the exit probability (sections \ref{ssec:estimates} and \ref{ssec:thermodynamic-limit}). The dual model can also be simulated much more efficiently, which allowed us to obtain definite results for network sizes up to $10^5$ (the previous works studied sizes up to $10^3$). Furthermore, the same arguments used to derive equation \ref{eq:bounds} can be used to obtain estimates about the exit probability from the transient, which we have done for system sizes up to $3.16\times 10^7$ (section \ref{sec:sims}). Our simulations show some small deviations from equation \ref{eq:q-continua}, but these only appear for network sizes around $10^5$ and bigger. Nevertheless, inequality \ref{eq:bounds} holds for all simulations done.

Finaly, we show (appendix \ref{ap:mean-field}) that the step function can be obtained as the exit probability for the $q$-voter model, using a mean field approach that neglects nearest neighbour correlations (in accordance to one of our previous works \cite{Timpanaro-2012}). We also show how the finite size effects should behave in this case and discuss how to reconcile this with the GUF (section \ref{sec:disc}). Because of this we believe that it is not correct to infer that fluctuations are neglectable only on the basis of equation \ref{eq:q-continua}.

\section{The one dimensional $q$-voter model}

The model studied in \cite{q-voter-Sznajd} (of which the model studied in \cite{q-voter-Galam, q-voter-Lambiotte, q-voter-Slanina} is a particular case) is defined in a linear chain with $N$ spins, that are either $+$ or $-$. The time evolution depends on an integer parameter, $q$, on the following way

\begin{itemize}
\item At each time step, $q$ neighbouring spins are chosen at random (these spins must be consecutive, $i, i+1, \ldots, i+q-1$).
\item If one of these spins has a state different from the others, then nothing happens.
\item Otherwise, two different versions of the model prescribe slightly different updates:
\begin{description}
\item[Both sides version] Spins $i-1$ and $i+q$ assume the same value as the spins in the group.
\item[Random version] Either $i-1$ or $i+q$, chosen at random, assumes the same value as the spins in the group.
\end{description}
\end{itemize}

\begin{center}
\begin{figure}[hbt]
\includegraphics[width = \columnwidth]{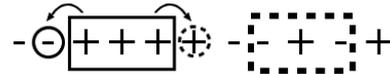}
\caption{Example of the model rules for $q=3$. The rectangles indicate 2 possible choices for the group of sites. The dashed rectangle represents a choice where we must go to the next time step without changing the state of the model. For the other rectangle, the circles represent the 2 sites that can be chosen to be convinced. The dashed circle represents a choice that doesn't change the state, while the full circle represents a site that should be flipped if it is the one that was chosen, changing the state of the system.}
\label{fig:exemplo-regra}
\end{figure}
\end{center}

In this work, we will only be concerned with the random version and we will always assume that the linear chain has periodic boundary conditions. The model with $q=1$ is the Ochrombel model and hence it is mathematicaly equivalent to the voter model, when using a linear chain. As the behaviour in this case is well known, we will focus in the case $q\geq 2$. An example of the rules in the case $q=3$ is given in figure \ref{fig:exemplo-regra}

\subsection{The dual model}
\label{ssec:dual}

\begin{center}
\begin{figure}[hbt]
\includegraphics[width = \columnwidth]{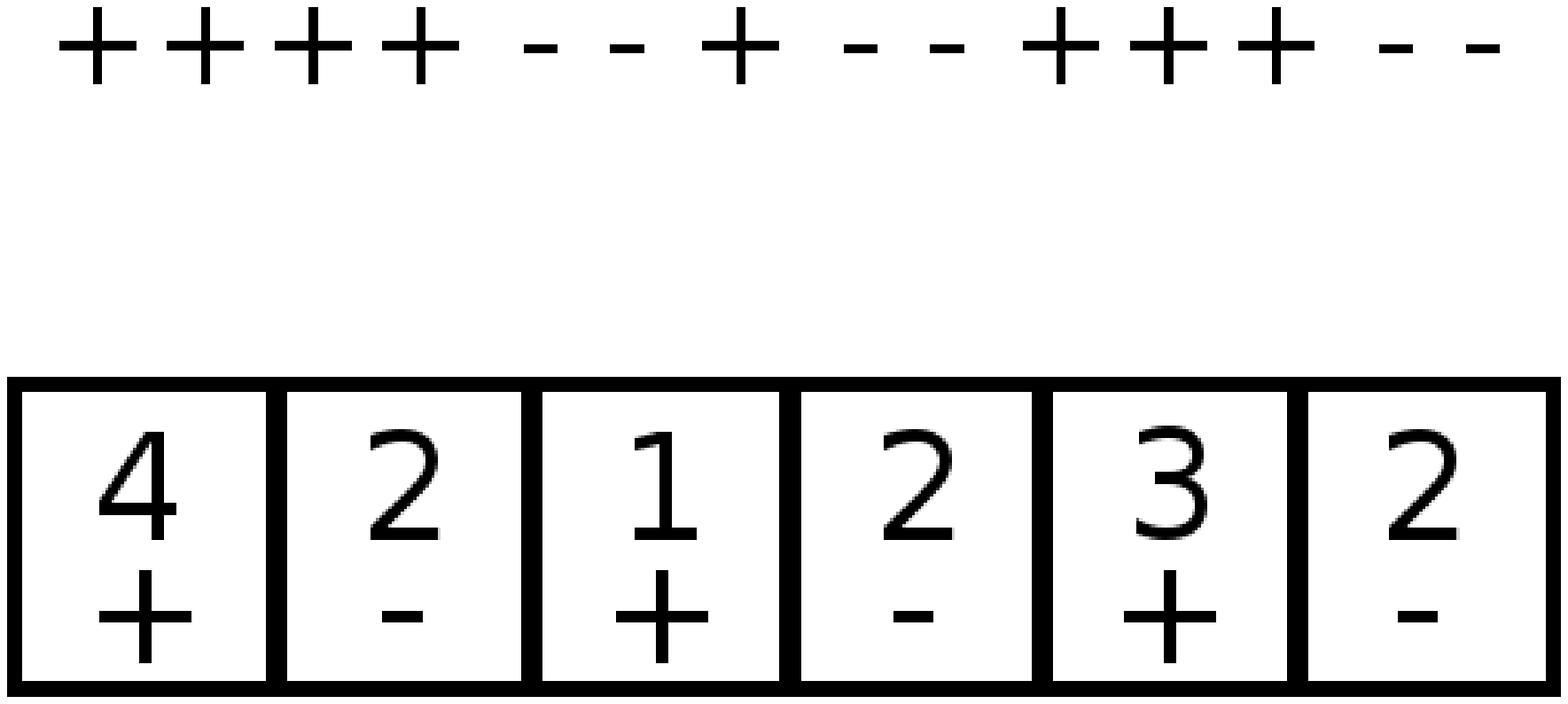}
\caption{Example of the mapping considered}
\label{fig:map}
\end{figure}
\end{center}

We now build a mapping between the one dimensional $q$-voter model and a dual model (meaning both models are mathematicaly equivalent), that as we will show has some similarities with the voter and Ochrombel models. Firstly, we note that the state of the original model can be described entirely by the sizes of contiguous groups of sites with the same spin (for example, 3 $+$ spins, 4 $-$ spins, 2 $+$ spins, 2 $-$ spins, etc.). We will then assign an index for each of these contiguous groups and use only their sizes and spins to keep track of the state of the model. A more detailed example of this mapping can be found in figure \ref{fig:map}.

Next, we note that the $q$ sites, that are chosen in the first step (of each iteration) of the original model, will have the same opinion if and only if they are all inside the same group. Moreover, the choice of these $q$ sites will change the state of the model if either they are the $q$ rightmost sites in the group and they choose to change the site to their right or if they are the $q$ leftmost sites and choose to change the site to their left (this can be seen in figure \ref{fig:exemplo-regra}). We can index all the possible choices for this group of $q$ sites and the neighbour to be convinced by $(i, \pm)$ in the following way:

\[
\left\{
\begin{array}{ll}
(i,+) \Rightarrow & \mbox{choose } i, \ldots, i+q-1 \mbox{ and convince } i+q \\
(i,-) \Rightarrow & \mbox{choose } i, \ldots, i+q-1 \mbox{ and convince } i-1
\end{array}
\right.
\]

\begin{center}
\begin{figure}[hbt]
\includegraphics[width = \columnwidth]{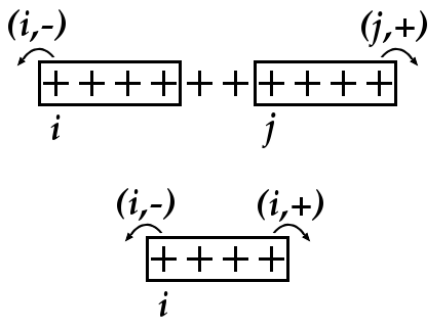}
\caption{Examples, for $q=4$, of choices of groups and neighbouring sites, that change the state of the model. Note that no matter the size of the contiguous group it has always 2 group-neighbour choices that change the state of the model, if it has at least $q$ members.}
\label{fig:scheme-probability}
\end{figure}
\end{center}

All of these choices have the same probability of happening. So, unless we are in a state where all sites agree, every contiguous group with at least $q$ sites sharing the same opinion has exactly one choice $(i,-)$ and one choice $(j,+)$, that changes the state of the model (the scheme in figure \ref{fig:scheme-probability} helps to ilustrate this). This means two things:

\begin{itemize}
\item If we skip all the updates that don't change the state of the model, the probability that the $q$ sites, chosen for the update that does change the state, are in a given group is 0 for all groups with less than $q$ sites and is the same for all the other groups, no matter what their sizes are.
\item The probability that the site to be changed is to the left of the group of contiguous sites is the same that it is to the right.
\end{itemize}

With these informations we build the dual model. We no longer have a chain of sites, but instead we have a chain with groups of sites, that can be merged if needed. We let $n_i$ be the number of sites in group $i$ and $s_i$ be their spin. The rules of the dual model are as follows:

\begin{itemize}
\item At each time step, choose a group $i$ at random, such that $n_i \geq q$.
\item Choose $r = \pm 1$ at random.
\item Decrease $n_{i+r}$ by 1 and increase $n_i$ by 1.
\item If this brings $n_{i+r}$ to 0, remove group $i+r$ and merge groups $i$ and $i+2r$. (this requires us to reindex all the groups and add together the sizes of groups $i$ and $i+2r$)
\end{itemize}

\begin{center}
\begin{figure}[hbt]
\includegraphics[width = \columnwidth]{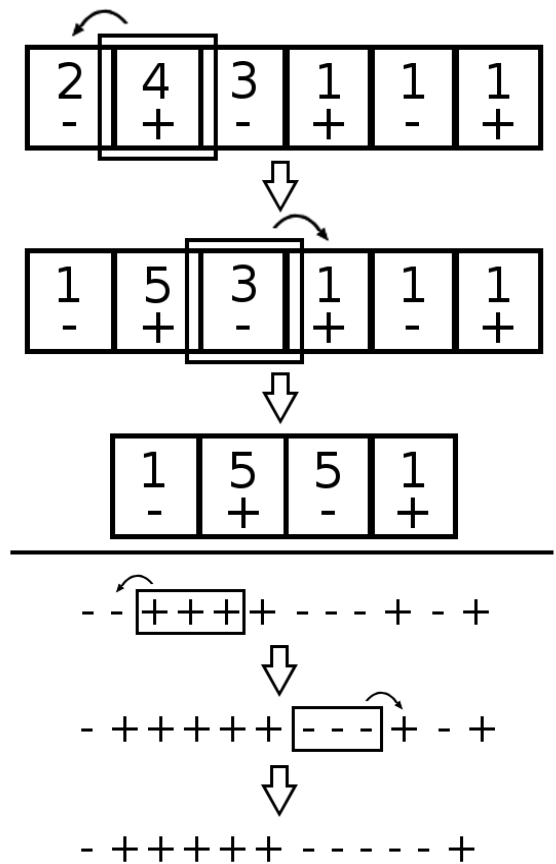}
\caption{Two updates in the dual model and their corresponding updates in the usual formulation of the $q$-voter model ($q = 3$ in the example). Note the merging rule being applied in the second update.}
\label{fig:dual-exemplo}
\end{figure}
\end{center}

Some examples of updates in the usual formulation, that change the system state and their corresponding updates in the dual formulation (including a case where merging is needed) can be seen in figure \ref{fig:dual-exemplo}.

The only real difference between this and the original model is that we are effectively skiping all the updates that don't change the state of the model. As such, the time variable must be updated carefully, but as we are only interested in the exit probability we don't need to worry about this \footnote{The probability that there will be $\tau$ steps of the original model, for one step of the dual model is $\nicefrac{G(1-\nicefrac{G}{N})^{\tau - 1}}{N}$, where $G$ is the number of groups such that $n\geq q$. As such, there are 2 possible ways to update the time variable. The first is to draw a real number $\xi\in[0,1[$ and to increase the time variable by $\left\lceil\nicefrac{\ln(\xi)}{\ln(1 - \nicefrac{G}{N})}\right\rceil$ steps. The second is a simplification where the time variable is increased by $\nicefrac{N}{G}$ (the average increase of the previous procedure).}. From an implementation point of view the chain of groups can be represented easily by a circular doubly linked list, while the groups with at least $q$ sites can be efficiently stored in an array.

\subsection{Connection with the voter model and estimates of the exit probability}
\label{ssec:estimates}

\begin{center}
\begin{figure}[hbt]
\includegraphics[width = \columnwidth]{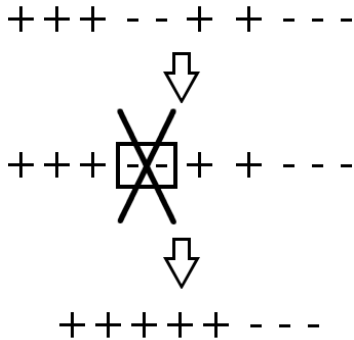}
\caption{An ilustration of our first hypothesis: The probability that $+$ becomes the dominant opinion in the final state is not smaller than the probability that it becomes the dominant one in the starting state, because we removed 2 $-$ spins from the network in the process.}
\label{fig:hip1}
\end{figure}
\end{center}

We now show that the voter model is related to a biased version of the dual model. Firstly, we recall that if we take $q=1$ we have the usual voter model. We then make the following hypothesis about the model:

\begin{itemize}
\item If at any point during the simulation we remove any site with opinion $-$ ($+$) (making the linear chain smaller), then this favours the opinion $+$ ($-$), in the sense that the probability that $+$ ($-$) becomes the dominant opinion does not become smaller.
\end{itemize}
An example of what this hypothesis means can be found in figure \ref{fig:hip1}.

\begin{center}
\begin{figure}[hbt]
\includegraphics[width = \columnwidth]{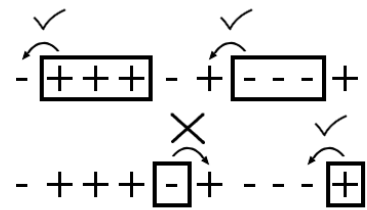}
\caption{An ilustration of our second hypothesis: In this case, we change the rules of the model, so that opinion $+$ doesn't need to form a group of size $q$ ($=3$) to convince sites with opinion $-$; while opinion $-$ needs to form such a group to convince sites with opinion $+$. Such a modification allows opinion $+$ to convince sites in the same situations that opinion $-$ can, including new situations in which opinion $-$ cannot and that are also not allowed in the unbiased version of the model. Because of this, we reason that this change doesn't make smaller the probability that $+$ becomes the dominant opinion.}
\label{fig:hip2}
\end{figure}
\end{center}

\begin{itemize}
\item If we allow that sites with opinion $+$ ($-$) be able to convince sites with opinion $-$ ($+$) even if they are not part of a group with $q$ agreeing agents, but we insist that sites with opinion $-$ ($+$) can only convince other sites if they are part of such group, then this favours opinion $+$ ($-$), in the same sense as before.
\end{itemize}
An ilustration of this hypothesis for $q=3$ can be found in figure \ref{fig:hip2}

These hypothesis are quite simple and intuitive. Moreover, even without a rigorous proof, they look rather sound. We use these hypothesis to build the following biased version of the $q$-voter model:

\begin{itemize}
\item At the beginning of the simulation, remove all sites that have opinion $-$ that are not in a group with at least $q$ members.
\item At each time step, choose a site $i$.
\item If $i$ has opinion $+$, then either $i+1$ or $i-1$, chosen at random, adopts opinion $+$.
\item If on the other hand, $i$ has opinion $-$, then check the sites $i+1, \ldots, i+q-1$. If they all have opinion $-$ , then either $i-1$ or $i+q$, chosen at random, adopts opinion $-$, but nothing happens otherwise.
\item Whenever a group containing sites with opinion $-$ drops below $q$ members, all the sites in the group are removed
\item Whenever 2 groups containing sites with opinion $-$ are merged (meaning they convinced all the sites with opinion $+$ that separated them), remove $q-1$ sites from the merged group.
\end{itemize}

\begin{center}
\begin{figure}[hbt]
\includegraphics[width = \columnwidth]{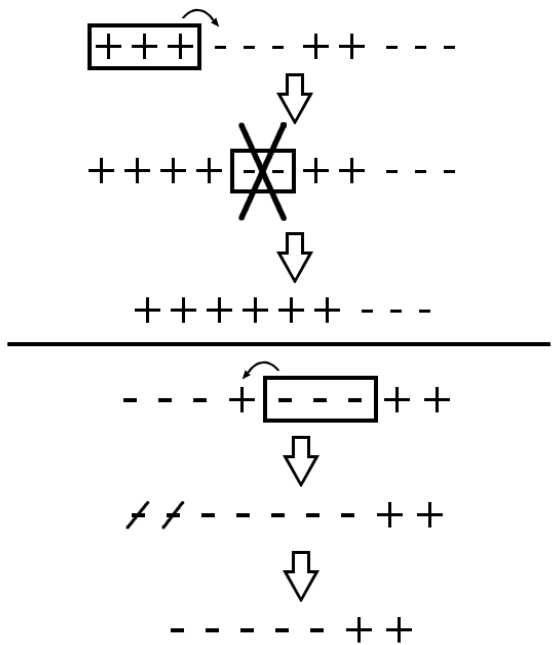}
\caption{The top example shows the removal of sites after the size of a group with opinion $-$ drops below $q$ ($=3$). The bottom example shows the removal of sites after 2 groups with opinion $-$ merge ($q=3$ again).}
\label{fig:exemplo-bias}
\end{figure}
\end{center}

In this version of the model we remove sites with opinion $-$, making the network smaller; at the beginning of the simulation, when we merge two groups of sites with opinion $-$ and when these groups become too small. We also require that sites with opinion $-$ be part of a group with $q$ agreeing sites in order to convince other sites, while we allow sites with opinion $+$ to convince other sites even if they are isolated. The change in the initial condition is identical to the example we gave in figure \ref{fig:hip1}. The other rules are ilustrated in figure \ref{fig:exemplo-bias}.

According to our two hypothesis, this means that this version is biased in favour of opinion $+$, in the sense that given the same initial conditions, the probability that $+$ becomes the dominant opinion in the $q$-voter model is at most the probability that it becomes the dominant one in the biased model. It is quite easy to make a version biased in favour of opinion $-$, meaning that these biased versions provide lower and upper bounds to the probability that an opinion becomes the dominant one, given an initial condition. It is interesting to see how these biased versions are translated in the language of the dual model:

\begin{itemize}
\item At the beginning of the simulation, remove all groups $i$ such that $s_i = -$ and $n_i < q$. Merge the remaining groups.
\item At each time step, choose a group $i$, such that either $n_i \geq q$ or $s_i = +$.
\item Choose $r = \pm 1$ at random.
\item Decrease $n_{i+r}$ by 1 and increase $n_i$ by 1.
\item If this brings $n_{i+r}$ to 0, remove group $i+r$ and merge groups $i$ and $i+2r$. If we also have $s_i = -$, decrease $n_i$ by $q-1$.
\item If on the other hand, this brings $n_{i+r}$ to $q-1$ and $s_{i+r} = -$, remove group $i+r$ and merge groups $i$ and $i+2r$.
\end{itemize}

Note that the groups with opinion $-$ can't have $n_i < q$, because the first rule eliminates these groups from the initial condition, while the last one guarantees that these groups are removed as soon as they are created by the dynamics. This means that the second rule is equivalent to choosing a group at random. Moreover, no matter the details of the stochastic evolution, as long as the final state has all sites with the same opinion, every group of sites $-$ that is not eliminated before the simulation starts will lose $q-1$ sites at some point, because the group will either be brought to $q-1$ sites and eliminated or merge with another group, losing $q-1$ sites during the process (actually, if the final state has all sites with opinion $-$, one of the groups didn't lose any sites. However, we can always remove $q-1$ sites of this final group and still get $-$ as the dominant opinion, so we can neglect this exception). This means that we can ``remove beforehand'' these $q-1$ sites and get the same dynamics. That is, the dual model biased in favour of opinion $+$ is equivalent to

\begin{center}
\begin{figure}[hbt]
\includegraphics[width = \columnwidth]{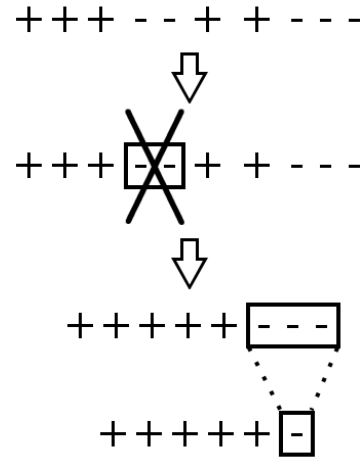}
\caption{The biased dual model can be formulated by modifying the initial condition and then following the rules of the usual voter model. Firstly we remove all groups with opinion $-$ and less than $q$ ($=3$) members. We merge the remaining groups and reduce all the remaining groups with opinion $-$ by $q-1$ members. This figure illustrates what is happening with the linear chain of spins during the modification of the initial condition and why this is a bias in favour of opinion $+$.}
\label{fig:bias_initial}
\end{figure}
\end{center}

\begin{itemize}
\item At the beginning of the simulation, remove all groups $i$ such that $s_i = -$ and $n_i < q$. Merge the remaining groups. For every remaining group $i$ such that $s_i = -$, decrease $n_i$ by $q-1$.
\item Follow the rules of the dual model for $q=1$ (that is, the voter model).
\end{itemize}
The effect in the initial condition is ilustrated in figure \ref{fig:bias_initial}.

This means that we can calculate the probability that $+$ ($-$) becomes the dominant opinion in the biased model (providing lower and upper bounds for this probability in the $q$-voter model) by biasing the initial condition and using the fact that the exit probability for the voter model is linear and independent of any correlations that exist in the initial condition. We can do this either for an initial condition drawn at random, which can be done analiticaly for an infinite system, or during the transient to get approximates for the exit probability much faster, allowing us to study larger system sizes numerically.

To get these estimates we need to measure the following quantities before biasing the model

\begin{itemize}
\item $N_s$: the total number of sites holding opinion $s$.
\item $\widetilde{N}_s$: the number of sites holding opinion $s$ that are in groups with at least $q$ elements.
\item $g_s$: the number of groups of sites holding opinion $s$, that have at least $q$ elements.
\end{itemize}
It's easy to see that after biasing the model in favour of opinion $+$, there are $N_{+}$ sites holding opinion $+$ and $\widetilde{N}_{-} - (q-1)g_{-}$ holding opinion $-$, while after biasing it in favour of opinion $-$ there are $\widetilde{N}_{+} - (q-1)g_{+}$ sites holding opinion $+$ and $N_{-}$ holding opinion $-$. To make the expressions simpler we define 
\[
\widehat{N}_s = \widetilde{N}_s - (q-1)g_s.
\]
It follows that the probability $E$ that the original $q$-voter model reaches the state where opinion $+$ is the dominant one can be bounded as 

\begin{equation}
\frac{\widehat{N}_{+}}{\widehat{N}_{+} + N_{-}} \leq E \leq \frac{N_{+}}{N_{+} + \widehat{N}_{-}}.
\label{eq:bound-geral}
\end{equation}
Finally, we can control the difference between the upper and lower bounds $\alpha \equiv E_{upper} - E_{lower}$ by stopping the simulation when

\begin{equation}
\frac{N_{+}N_{-} - \widehat{N}_{+}\widehat{N}_{-}}{(N_{+} + \widehat{N}_{-})(N_{-} + \widehat{N}_{+})} \leq  \alpha ,
\label{eq:criterio-de-parada}
\end{equation}
for an acceptable value of $\alpha$.

\subsection{Estimates for a system in the infinite size limit}
\label{ssec:thermodynamic-limit}

To get estimates for a system in the infinite size limit we need to find the values of

\[
\frac{N_s}{N}, \frac{g_s}{N} \mbox{ and } \frac{\widetilde{N}_s}{N}
\]
as the lenght of the chain, $N$, goes to infinity. Suppose then that we choose opinion $+$ with probability $\rho$ while drawing the initial condition. We can regard the initial condition as a sequence of groups with spins alternating between $+$ and $-$. Also, the probability of a $+$ group being drawn with size $k$ is $\rho^{k-1}(1-\rho)$, while this probability is $\rho(1-\rho)^{k-1}$ for a group with spin $-$. So if we take $\rho\neq 0,1$ we have after drawing $G$ groups

\begin{equation}
N\rho = \langle N_{+}\rangle  = G\sum_{k=1}^{\infty} k\rho^{k-1}(1-\rho),
\end{equation}
\begin{equation}
\langle g_{+}\rangle  = G\sum_{k=q}^{\infty} \rho^{k-1}(1-\rho) = G\rho^{q-1}\mbox{ and}
\end{equation}
\begin{equation}
\langle \widetilde{N}_{+}\rangle  = G\sum_{k=q}^{\infty} k\rho^{k-1}(1-\rho).
\end{equation}
After some algebraic manipulations this leads to

\[
\langle \widetilde{N}_{+}\rangle  = \rho^{q-1}(\langle N_{+}\rangle  + G(q-1)) \Rightarrow
\]\[
\langle \widetilde{N}_{+}\rangle  = N\rho^{q} + (q-1)\langle g_{+}\rangle \Rightarrow
\]
\begin{equation}
\langle \widehat{N}_{+}\rangle  = \langle \widetilde{N}_{+}\rangle  - (q-1)\langle g_{+}\rangle  = N\rho^q.
\end{equation}
Exchanging $\rho$ for $1-\rho$ we can get $\langle N_{-}\rangle  = N(1-\rho)$ and $\langle \widehat{N}_{-}\rangle = N(1-\rho)^q$. Substituting in equation \ref{eq:bound-geral} leads to

\begin{equation}
\frac{\rho^q}{\rho^q + (1 - \rho)} \leq E \leq \frac{\rho}{\rho + (1-\rho)^q}.
\label{eq:bound-inicial}
\end{equation}

\section{Simulation Results}
\label{sec:sims}

\begin{figure}[hbt]
\includegraphics[width = 0.95\columnwidth]{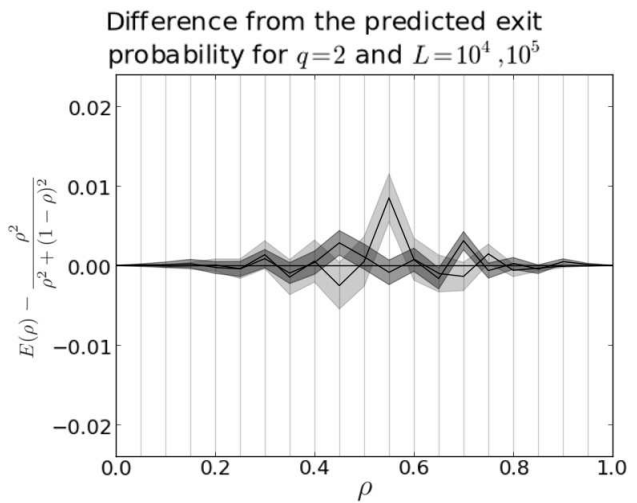}
\caption{Difference between the exit probability measured by waiting the system go to a stationary state and the one predicted by equation \ref{eq:q-continua} (using $q=2$). The system sizes are $L=10^4$ (dark gray) and $L=10^5$ (light gray). The width of the band represents the statistical deviation of the measures taken.}
\label{fig:diffs-stat}
\end{figure}

\begin{figure}[hbt]
\includegraphics[width = 0.95\columnwidth]{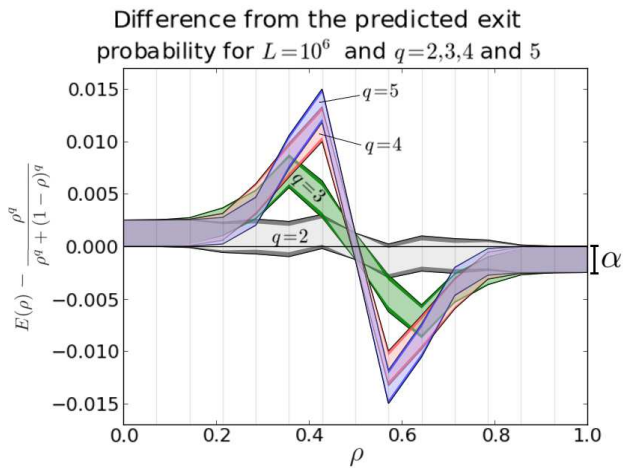}
\caption{(Color online) Difference between the exit probability, estimated from the transient using equation \ref{eq:bound-geral}, and the one predicted by equation \ref{eq:q-continua} for a system size equal to $L=10^6$ and $q=2,3,4,5$. The lighter part of each band represents the margin $\alpha$ between the upper and the lower bounds and the darker part represents the statistical deviation of the two bounds. The evaluated points were in the range $0 < \rho < \nicefrac{1}{2}$ and the rest explores the symmetry $E(\rho) + E(1-\rho) = 1$.}
\label{fig:diffs-6}
\end{figure}

\begin{figure}[hbt]
\includegraphics[width = 0.95\columnwidth]{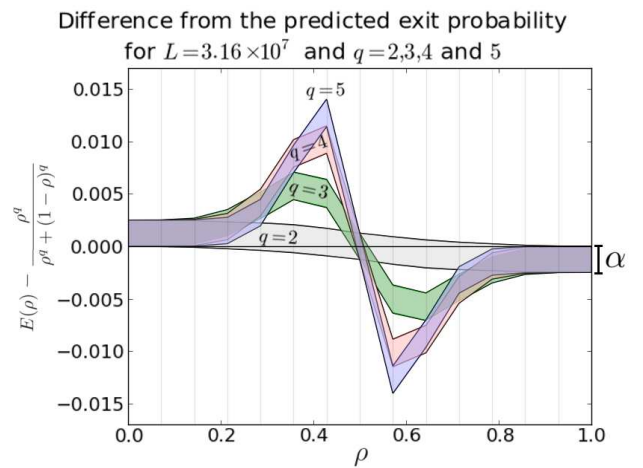}
\caption{(Color online) Difference between the exit probability, estimated from the transient using equation \ref{eq:bound-geral}, and the one predicted by equation \ref{eq:q-continua} for a system size equal to $L=3.16\times 10^7$ and $q=2,3,4,5$. Each band represents the margin $\alpha$ between the upper and the lower bounds for each value of $q$ (the statistical deviation of the bounds is too small to appear). This graph is undistinguishable from the graph obtained for a system size equal to $L=10^7$. The evaluated points were in the range $0 < \rho < \nicefrac{1}{2}$ and the rest explores the symmetry $E(\rho) + E(1-\rho) = 1$.}
\label{fig:diffs-7-5}
\end{figure}

We made simulations for system sizes $L = 10^4$ and $L=10^5$ using the dual model with $q=2$ and waiting until a consensus state was reached. We also did simulations for system sizes $L=10^6$, $L=3.16\times 10^6$, $L=10^7$ and $L=3.16\times 10^7$, using the dual model for $q=2, 3, 4$ and 5, stopping when the difference between the upper and lower bounds was $\alpha = 2.5\times 10^{-3}$ (see equations \ref{eq:bound-geral} and \ref{eq:criterio-de-parada}). The results of these simulations, compared with the exit probabilities proposed in \cite{q-voter-Sznajd} are in figures \ref{fig:diffs-stat}, \ref{fig:diffs-6} and \ref{fig:diffs-7-5}.

Our results for the simulations where we waited until a consensus was reached show that for $q=2$, using $L=10^4$ and $L=10^5$, the exit probability couldn't be distinguished from
\begin{equation}
E = \frac{\rho^2}{\rho^2 + (1 - \rho)^2},
\label{eq:proposta-q-2}
\end{equation}
as proposed in \cite{q-voter-Sznajd} and \cite{q-voter-Lambiotte}. The same thing was observed for $q=2$ in our estimates for system sizes bigger than $L=10^6$. But for $q>2$, the exit probability we obtained was consistently higher for $\rho < \nicefrac{1}{2}$ and consistently smaller for $\rho > \nicefrac{1}{2}$, when compared with the formula in equation \ref{eq:proposta-q-2}. The absolute difference was as high as 0.015 for $q=5$ and $\rho$ close to 0.4. It's also of note that the discrepancies increased with $q$, but in the opposite way that would be expected if the exit probability were to be a step function.

%
%

\section{Conclusions}

On this work, we have studied the exit probability of a version of the one dimensional $q$-voter model in one dimension, previously studied in \cite{q-voter-Slanina, q-voter-Lambiotte, q-voter-Galam, q-voter-Sznajd}. In these previous works, network sizes up to $10^3$ sites were studied. We have presented here an algorithm that makes it feasible to study much larger network sizes (we studied this way network sizes up to $10^5$ sites). We have also developed a way to make estimates of the exit probability from the transient, based on some simple assumptions (simulations in this case were made for up to $3.16\times 10^7$ sites).

These new simulations are able to shed light in some of the controversies that have arised in this model. Our results for the exit probability with $q=2$ are undistinguishable from

\[
E(\rho) = \frac{\rho^2}{\rho^2 + (1-\rho)^2}
\]
for network sizes up to $3.16\times 10^7$ sites, as proposed in \cite{q-voter-Slanina, q-voter-Lambiotte}, based on simulations for small sizes and on the Kirkwood approximation, and in contrast with the suggestion made in \cite{q-voter-Galam} that the exit probability should be a step function, as would be expected from the application of the GUF. The way to reconcile the GUF with these results is to notice that this approach makes no reference to any social network, so if the exit probability is different for 2 networks, then the GUF can't give the correct precision for both these cases. As we show in the appendix \ref{ap:mean-field}, the GUF exit probability coincides with the exit probability in a complete graph, which corresponds to a mean-field neglecting pair correlations (and is in accordance with our previous work about the mean-field of this model, in a more general context \cite{Timpanaro-2012}). For this same reason, the fact that the exit probability for $q=2$ can be predicted by the Kirkwood approximation (which is a mean-field approach neglecting correlations beyond pairs) is not sufficient to conclude that fluctuations are neglectable in this system, as a different exit probability can also be deduced from a different mean-field approach.

Finaly, the discrepancies found in the cases $q>2$ show that the arguments given in \cite{q-voter-Sznajd} for the exit probability

\[
E(\rho) = \frac{\rho^q}{\rho^q + (1-\rho)^q}
\]
do not provide a completely acurate picture.


\appendix


\section{The mean field exit probability}
\label{ap:mean-field}

We now investigate the mean field exit probability, arguing that it must be a step function. Consider the mean-field version of the $q$-voter model with 2 states (up and down) and $N$ agents. The state of the model is completely determined by the number $n$ of agents holding opinion up. Accordingly, there must be a function $E_N(n)$ called the exit probability that gives the probability that starting in the state with $n$ agents up and $N-n$ agents down we end in the absorbing state where all agents have opinion up.

As this function has no explicit dependency with time, the following equation must hold:

\[
E_N(n) = p_{-}(N,n)E_N(n-1) + p_0(N,n)E_N(n) +
\]
\begin{equation}
+ p_{+}(N,n)E_N(n+1),
\label{eq:time-displacement-consistency}
\end{equation}
where $p_{-}(N,n), p_0(N,n), p_{+}(N,n)$ are the probabilities that after one iteration we have respectively the transitions $n\rightarrow n-1$, $n\rightarrow n$ and $n\rightarrow n+1$, when we have a total of $N$ agents (hence $p_{-} + p_0 + p_{+} = 1$). The reasoning is that $E_N(n)$ is the probability that all spins end in the up state and as such all the possible ways for this to happen must be accounted. As we are in the mean-field approximation, there are no spatial correlations, and so after one iteration we can still use the same functional form for the exit probability. After the first iteration, either an up spin flipped with probability $p_{-}$, a down spin flipped with probability $p_{+}$ or nothing happened. The recurrence relation (\ref{eq:time-displacement-consistency}) expresses then the fact that the probability can be calculated consistently either in a given iteration, or right after that iteration. For the $q$-voter model we have

\begin{equation}
p_{-}(N,n) = \frac{(N-n)(N-n-1)\ldots (N-n-q+1)n}{N(N-1)\ldots (N-q+1)(N-q)}
\label{eq:p-minus}
\end{equation}
and

\begin{equation}
p_{+}(N,n) = \frac{n(n-1)\ldots (n-q+1)(N-n)}{N(N-1)\ldots (N-q+1)(N-q)}.
\label{eq:p-plus}
\end{equation}

Equation \ref{eq:time-displacement-consistency} can be rewritten defining $\Delta E_N(n) = E_N(n+1) - E_N(n)$ as

\begin{equation}
\Delta E_N(n+1) = \Delta E_N(n).\frac{p_{-}(N,n+1)}{p_{+}(N,n+1)}.
\label{eq:cm-diff}
\end{equation}

Moreover, in the $q$-voter model, if $N\geq 2q-1$, then the exit probability must obey $E_N(r) = 0$ for $r = 0, \ldots, q-1$ and $E_N(r) = 1$ for $r = N-q+1,\ldots, N$. Hence, as equation \ref{eq:cm-diff} is a linear recurrence relation, we can write the solution as

\begin{equation}
\Delta E_N(n) = c \prod_{s=q}^{n}\frac{p_{-}(N,s)}{p_{+}(N,s)},
\label{eq:cm-main}
\end{equation}
where $c$ works as a normalization constant (actually, $c = E_N(q) = \Delta E_N(q-1)$) determined only by the requirement that $E_N(N) = 1$. Substituting $p_{-}$ and $p_{+}$ in equation \ref{eq:cm-main} we get


\[
\Delta E_N(n) = c \prod_{r=1}^{q-1}\prod_{s=q}^{n}\frac{N-s-r}{s-r} \Rightarrow
\]




\begin{equation}
\Delta E_N(n) = c \prod_{r=1}^{q-1}\frac{(N-r-q)!(q-r-1)!}{(n-r)!(N-n-r-1)!} = 
\label{eq:cm-factorial-decomposition}
\end{equation}
\[
= c\left(\prod_{r=1}^{q-1}\binom{N-2r-1}{q-r-1}\right)^{-1} \left(\prod_{r=1}^{q-1}\binom{N-2r-1}{n-r}\right)
\Rightarrow
\]
\begin{equation}
\Delta E_N(n) = c.C_{N,q}\prod_{r=1}^{q-1}\binom{N-2r-1}{n-r},
\label{eq:cm-binomial-decomposition}
\end{equation}
where the $C_{N,q}$ are defined as

\begin{equation}
C_{N,q} = \left(\prod_{r=1}^{q-1}\binom{N-2r-1}{q-r-1}\right)^{-1}
\label{eq:cnq-definition}
\end{equation}
and hence, they don't depend on $n$ and can be absorbed in the normalization constant.

Finaly, for large $N$ we can make the approximation
\begin{equation}
E'(\rho) = \frac{A_{N,q}}{B(N\rho ,N(1-\rho))^{q-1}},
\label{eq:mf-continuous}
\end{equation}
where $E(\rho) = E_N(N\rho)$, $B(x,y)$ is the beta function and $A_{N,q}$ is a normalization constant (we get the power $q-1$ because $\Delta E_N$ is the product of $q-1$ binomials).

The function in equation \ref{eq:mf-continuous} is symmetric about $\rho = \nicefrac{1}{2}$ and has a peak for this value if $q>1$. The resulting $E(\rho)$ is a sigmoid and we can check how it behaves in the limit $N\rightarrow\infty$ by checking the width of the peak in $E'(\rho)$. When $N\rho, N(1-\rho)\rightarrow\infty$ we can use the Stirling approximation to get

\begin{equation}
E'(\rho) \approx \frac{A_{N,q}}{(\rho^{\rho}(1-\rho)^{(1-\rho)})^{N(q-1)}}.
\label{eq:mf-approx}
\end{equation}

We can check the behaviour of the width $\gamma$ at half maximum height in equation \ref{eq:mf-approx} by solving

\begin{equation}
(1+2\gamma)^{(1+2\gamma)}(1-2\gamma)^{(1-2\gamma)} = 4^{\nicefrac{1}{N(q-1)}},
\label{eq:half-width-half-maximum}
\end{equation}
that is asymptoticaly (when $N\rightarrow \infty$, $\gamma\rightarrow 0$)

\[
(1+2\gamma(1+2\gamma))(1-2\gamma(1-2\gamma)) = 1 + \frac{\log 4}{N(q-1)} \Rightarrow
\]\[
1 + 4\gamma^2 = 1 + \frac{\log 4}{N(q-1)} \Rightarrow
\]
\begin{equation}
\gamma\approx\frac{1}{\sqrt{N(q-1)}}.
\label{eq:width-asympt}
\end{equation}

This shows that for $q\geq 2$ the exit probability $E(\rho)$ tends to a step function and the finite size effects observed are straight forward. The classical result for the voter model ($q=1$) follows from equation \ref{eq:mf-continuous}, that reduces to $E'(\rho)=\mathrm{constant}$ and hence $E(\rho)=\rho$ because of the boundary conditions for $\rho = 0, 1$.

\bibliography{andre}

\begin{thebibliography}{10}

\bibitem{q-votante-def}
C.~Castellano, R.~Pastor-Satorras, and M.~A.~Mu\~ noz, ``Nonlinear q-voter
  model,'' {\em Physical Review E}, vol.~80, no.~4, p.~041129, 2009.

\bibitem{sznajd-def}
K.~Sznajd-Weron and J.~Sznajd, ``Opinion evolution in closed community,'' {\em
  International Journal of Modern Physics C}, vol.~11, no.~6, pp.~1157--1165,
  2000.

\bibitem{q-voter-Slanina}
F.~Slanina, K.~Sznajd-Weron, and P.~{Przyby\l a}, ``Some new results on
  one-dimensional outflow dynamics,'' {\em Europhysics Letters}, vol.~82,
  p.~18006, 2008.

\bibitem{q-voter-Lambiotte}
R.~Lambiotte and S.~Redner, ``Dynamics of non-conservative voters,'' {\em
  Europhysics Letters}, vol.~82, p.~18007, 2008.

\bibitem{q-voter-Galam}
S.~Galam and A.~C.~R. Martins, ``Pitfalls driven by the sole use of local
  updates,'' {\em Europhysics Letters}, vol.~95, p.~48005, 2011.

\bibitem{kirkwood-approximation}
J.~G. Kirkwood, ``The radial distribution function in liquids,'' {\em Journal
  of Chemical Physics}, vol.~3, p.~300, 1935.

\bibitem{Galam-GUF}
S.~Galam, ``Local dynamics vs social mechanisms: A unifying frame,'' {\em
  Europhysics Letters}, vol.~70, no.~6, p.~705, 2005.

\bibitem{q-voter-Sznajd}
P.~{Przyby\l a}, K.~Sznajd-Weron, and M.~Tabiszewski, ``Exit probability in a
  one-dimensional nonlinear q-voter model,'' {\em Physical Review E}, vol.~84,
  p.~031117, 2011.

\bibitem{Timpanaro-2012}
A.~M. Timpanaro and C.~P.~C. do~Prado, ``Connections between the sznajd model
  with general confidence rules and graph theory,'' {\em Physical Review E},
  vol.~86, p.~046109, 2012.

\bibitem{Note1}
The probability that there will be $\tau $ steps of the original model, for one
  step of the dual model is $\protect \nicefrac {G(1-\protect \nicefrac
  {G}{N})^{\tau - 1}}{N}$, where $G$ is the number of groups such that $n\geq
  q$. As such, there are 2 possible ways to update the time variable. The first
  is to draw a real number $\xi \in [0,1[$ and to increase the time variable by
  $\left\lceil\nicefrac{\ln(\xi)}{\ln(1 - \nicefrac{G}{N})}\right\rceil$. The second is a simplification where the time variable is increased by $\nicefrac{N}{G}$ (the average increase of the previous procedure).

\end{thebibliography}
\bibliographystyle{ieeetr}

\end{document}